\documentclass[prl,twocolumn,showpacs,superscriptaddress,preprintnumbers,amsmath,amssymb,floatfix]{revtex4}

\usepackage{bm}
\usepackage{graphicx}
\usepackage{amsbsy}
\usepackage{amsmath}
\usepackage{amsfonts}
\usepackage{amsthm}
\usepackage{color}
\usepackage{mathrsfs}
\usepackage{graphicx}
\usepackage{dcolumn}
\usepackage{bm}
\def\ket#1{|\,#1\,\rangle}
\newcommand{\bwt}{\begin{widetext}}
\newcommand{\ewt}{\end{widetext}}
\newcommand{\bea}{\begin{eqnarray}}
\newcommand{\eea}{\end{eqnarray}}

\begin{document}

\title{Optical quantum memory with generalized time-reversible atom-light interactions}

\author{S.A. Moiseev}
\affiliation{Institute for Quantum Information Science, University of Calgary,
        Calgary, Alberta, Canada T2N 1N4}
\affiliation{Kazan Physical-Technical Institute of the Russian Academy of Sciences,
10/7 Sibirsky Trakt, Kazan, 420029, Russia}

\author{W. Tittel}
\affiliation{Institute for Quantum Information Science, University of Calgary,
        Calgary, Alberta, Canada T2N 1N4}


\begin{abstract}

We examine a quantum memory scheme based on controllable dephasing of atomic coherence of a non-resonant, inhomogeneously broadened Raman transition. We show that it generalizes the physical conditions for time-reversible interaction between light and atomic ensembles from weak to strong fields and from linear to non-linear interactions. We also develop a unified framework for different realizations exploiting either controlled reversible inhomogeneous broadening or atomic frequency combs, and discuss new aspects related to storage and manipulation of quantum states.

\end{abstract}

\pacs{03.67.-a, 42.50.Ct, 42.50.Md}


\maketitle


The study of time-reversible evolution of physical systems was central in the development of thermodynamics and statistical mechanics \cite{Zeh1992}, and CPT symmetry \cite{Gibson1976}. Furthermore, reversible interaction underpins reversible transfer of quantum states between light and atoms, i.e. quantum memory (QM), which is key for quantum repeaters \cite{Briegel1998} and all-optical quantum computing \cite{Kok2007}.

Approaches to QM exploit atoms in cavities \cite{Cirac1997}, non-resonant Raman transitions \cite{Kuzmich2000,Julsgaard2004}, electromagnetically induced transparency \cite{Fleischhauer2000,Eisaman2005,Chaneliere2005,Novikova2007,Choi2008,Appel2008,Honda2008}, or photon-echo techniques
\cite{Moiseev2001,Nilsson2005,Kraus2006,Alexander2006,Tittel2008,Hetet2008a,Afzelius2009,Riedmatten2008}. The latter is of particular interest in the present context, as the equations of motion include a hidden time-reversal symmetry \cite{Kraus2006}. It has, however, so far only been discussed for weak light fields.
Here we generalize the physical conditions allowing for time-reversal symmetry in the mapping of quantum states between light and atomic ensembles to fields of arbitrary strength and non-linear interactions. The scheme exploits reversible dephasing of atomic coherence of a non-resonant, inhomogeneously broadened Raman transition (Raman Echo Quantum Memory or REQM). It has been proposed in \cite{LasPhys2008}, further developed in \cite{Nunn2008,Gouet2009}, and first demonstrated in \cite{Hetet2008b}. We also compare the conditions for time-reversibility for storage of strong fields with those for weak fields, which naturally leads to a unified framework for realizations of REQM based on controlled reversible inhomogeneous broadening (CRIB) \cite{Kraus2006} or atomic frequency combs (AFC) \cite{Afzelius2009}.
Our findings shed new light on time-reversibility in the interaction between light and inhomogeneously broadened atomic ensembles, and also pave the road to storage of macroscopic light fields in nano-sized atomic media. Furthermore, we discuss with the example of frequency conversion how REQM enables controlled manipulation of quantum light fields.

\begin{figure}
  \includegraphics[width=0.4\textwidth,height=0.2\textwidth]{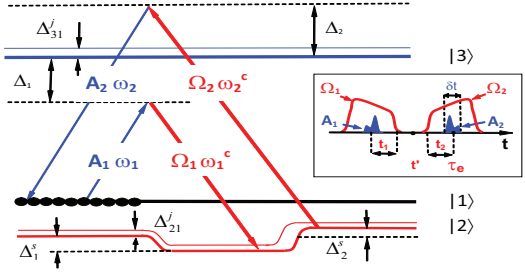}
  \caption{Energy level diagram showing atomic transitions, probe and echo fields
($A_{\nu}$) with carrier frequencies $\omega_{\nu}$, and writing and reading fields with Rabi frequencies
$\Omega_{\nu}$ and carrier frequencies
$\omega_{\nu}^c$. $\nu=1,2$ denotes storage and recall, resp. Also depicted is a temporal diagram of all light fields. }
  \label{RamFig1}
\end{figure}

\emph{The scheme:}
Energy and temporal diagrams of the interaction scheme are depicted in
Fig.\ref{RamFig1}. At time t=0 the probe light field $\hat{E}_{1}(t,z)$
with duration $\delta t_1$, carrier frequency $\omega_1$ and spectral width
$\delta\omega_1= \delta t_1^{-1}$ enters the medium with three-level
atoms (labeled by $j$) prepared in the long-lived level $\ket{1}=\prod_{j=1}^{N} \ket{1}_{j}$ along
the $+z$ direction. The atoms are simultaneously exposed to an
intense control (writing) field propagating along wavevector $\vec{K}_{1}$ with carrier frequency $\omega_1^c$ and
Rabi frequency $\tilde{\Omega}_{1}(t)$. It is reduced to zero 
after absorption of the probe field. The probe and writing fields
are assumed to be in Raman resonance
$\omega_{1}-\omega_{1}^c\approx\omega_{21}$ with sufficiently large
spectral detuning $\Delta_{1}=\omega_{31}-\omega_{1}$ from the
 $1\leftrightarrow 3$ transition.

We take the $1\leftrightarrow 2$ and $1\leftrightarrow 3$ transitions to
feature inhomogeneous broadenings (IB), resulting in a total detuning of each atom $j$
from the center of the Raman transition given by
$\Delta_{R,tot}^{j}(t)\cong
\Delta_{21,tot}^j+\Delta_{31,tot}^j f_1(t)$
with $\Delta_{k1,tot}^j=\Delta_{k1,nat}^j+\Delta_{k1,cont}^j\equiv\Delta_{k1}^j$, $(k=2,3)$.
The indexes refer to total and natural detuning, and possibly detuning
induced in a controlled way using e.g. external electric fields. We take the IB of the Raman transition to be large enough to absorb all spectral components of the probe field, and assume the natural IB on the $1\leftrightarrow 2$ transition to be negligible (as is usually the case for hyperfine ground states of rare-earth-ion doped crystals \cite{Tittel2008}). $f_1(t)$ is related to the Rabi frequency of the intense light field, $f_1 (t) =|\Omega_{1}(t)|^2/\Delta_{1}^2$, as well as to the Stark shift $\Delta_1^s$ induced by the control field, $f_1(t)=\Delta_1^s/\Delta_1$. It can exceed unity by orders of magnitude.

The atom-light interaction leads to excitation of atomic coherence, which rapidly decays due to IB. To retrieve the stored field, we apply a phase (or mode) matching operation, and launch at time $t>t^{\prime}$ a second control (reading) pulse propagating roughly in $-z$ direction and with wave vector, carrier and Rabi frequencies $\vec{K_2}$, $\omega_2^c$ and $\tilde{\Omega}_{2} (t)$, resp. (see Fig. 1). In addition, we either actively invert the Raman broadening, as in CRIB \cite{Kraus2006} (now RECRIB), or, in the case of a generalization of AFC \cite{Afzelius2009} (now REAFC), simply wait until the atomic coherence automatically rephases. The probe field, at frequency $\omega_2\approx\omega_{21}+\omega_2^c$, will then be re-emitted  at time $\tau_e$ as an echo in the backward ($-z$) direction. In the following we will limit our discussion to the case where the spectral width and duration of the echo are equal to the ones of the probe field: $\delta \omega_2=\delta \omega_1\equiv \delta \omega$, $\delta t_2=\delta t_1\equiv\delta t$.

We emphasize that the memory's storage bandwidth, which depends on the IB of the Raman transition, does not only rely on material properties but also on the control field via $f_1$ \cite{comment0}. Also, the direct Raman transfer allows using atomic materials with short optical coherence times \cite{Hetet2008b,Gouet2009}. These two features result in a larger choice of materials compared to photon-echo quantum memory protocols without direct Raman transfer.

\emph{Basic equations:}
We describe the interaction between the three-level atoms, probe and echo fields $\hat{E}_{\nu}(z)=\hat{A}_{\nu}(z)\exp \{i k_{\nu}z\}$, and control fields $\tilde{\Omega} _{\nu} (t,\vec{r})=\Omega _{\nu} (t )$ $\exp \{ - i(\omega _{\nu}^c t-\vec {K}_\nu \vec {r})\}$ using the Hamiltonian $\hat {H} = \hat {H}_a +\hat {H}_f + \hat {V}_{a-f} + \hat {V}_c$. $\nu=1,2$ labels storage and recall, and the indices $a,f,a-f$ and $c$ denote the Hamiltonian for the atoms, quantum light fields, and interaction between quantum and classical light fields and atoms, resp.:
\begin{eqnarray}
\begin{array}{ccc}
\hat {H}_a=\sum\limits_{j = 1}^N \sum\limits_{n = 1}^3 E_{nn}^j \hat{P}_{nn}^j,\nonumber\\
\hat {H}_f = \hbar \sum\limits_{{\nu} = 1}^2 {\int
{dz} \hat {A}_{\nu}^ + (z)
[\omega_{\nu}+i( - 1)^{\nu} \textstyle{v_{\nu}  }
\textstyle{\partial \over
{\partial z}}]\hat {A}_{\nu} (z)},\nonumber\\
\hat {V}_{a - f} = - \hbar \sum\limits_{j = 1}^N \sum\limits_{{\nu} = 1}^2
[g_{\nu} \hat {A}_{\nu} (z_j )
\exp\{i k_{\nu} z_{j}\}
\hat {P}_{31}^j + h.c. ],\nonumber\\
\hat {V}_c = - \hbar \sum\limits_{j = 1}^N \sum\limits_{{\nu} = 1}^2
[\Omega _{\nu} (t)\exp \{ - i(\omega _{\nu}^c t - \vec{K}_{\nu} \vec{r}_j)\} \hat {P}_{32}^j +
h.c.].\nonumber\\
\end{array}
\end{eqnarray}

\noindent
$\left[\hat{A}_{\nu '} (z'),\hat{A}_{\nu}^ + (z)\right]=\delta_{\nu ',\nu}\delta (z'-z),$
$k_{\nu}=(-1)^{\nu+1}$ $\frac{\omega_{\nu}n_{\nu}}{c}$,
$n_\nu$ and $v_{\nu} = {\partial \omega }/{\partial k}|_{\omega =
\omega _{\nu} }$ are the refractive indexes and group velocities for the probe and echo fields in the absence of interaction with the three-level atoms,
and $g_{\nu}$ is the photon-atom coupling constant.

In the following we use the Heisenberg picture and derive the equations of motion for the slowly varying operators for the light fields ($\hat{A}_{\nu,o}$) and atomic coherences ($\hat{R}_{mn,\nu}^j$) between states $m$ and $n$: $\hat{A}_{\nu}(z,t)=\hat{A}_{\nu,o}(z,t)\exp\{-i\omega_{\nu}t\})$,
$\hat{P}_{12}^j(t)=\hat {R}_{12,\nu}^j (t)$ $\exp\{ i\varphi _{\nu}(\vec{r},z_j)-i(\omega_{\nu}-\omega_{\nu}^c)(t+(-1)^{\nu}n_{\nu} z_j/c)\}$, $\hat {P}_{13}^j (t) = \hat {R}_{13,{\nu}}^j (t)$ $\exp \{ - i\omega _{\nu} (t + ( -1)^{\nu}n_{\nu} z_j / c)\}$,
$\hat {P}_{32}^j (t) = \hat {R}_{32,\nu}^j (t)\exp \{i\varphi _{\nu} (\vec{r},z_j)+i\omega _{\nu}^c (t + ( - 1)^{\nu} n_{\nu} z_j / c)\}$, $\hat {P}_{mm}^j(t )=\hat {R}_{mm,\nu}^j (t)$, and
$\varphi _{\nu} (\vec{r},z_j) = -((-1)^{\nu}n_{\nu}\omega _{\nu}^c z_j/c + \vec
{K}_{\nu} \vec {r})$. We also assume $|\Delta _{\nu}+ \Delta_{31,\nu}^j|$ $\gg\delta\omega$ (with arbitrary $\Delta _{\nu}/\Delta_{31,\nu}^j$), $|\Delta_{21,{\nu}}^j|\ll|\Delta_{31,{\nu}}^j|$, $|\Delta_{32,{\nu}}^j|$, and
$\frac{1}{\Delta _{\nu} + \Delta
_{32,{\nu}}^j }\cong\frac{1}{\Delta _{\nu} + \Delta
_{31,{\nu}}^j }$. Being an obvious condition for time-reversibility, we ignore all atomic decay as well as irreversible atomic dephasing, and assume that the probe field is completely absorbed in the atomic media.
To simplify the expressions, we use a coordinate system moving with the probe, or echo fields, resp. For absorption
($t<t^{\prime}$, ${\nu}=1$), we use $\tau_{1}=t-Z/{v_1}$, and for retrieval
($t>t^{\prime}$, ${\nu}=2$), we have $\tau_{2}=t+Z/{v_2}-\tau_e$,
where $Z=z$.
Finally, we use new quantum fields $\hat {\zeta}_{\nu} (\tau_{\nu},Z) =-(-1)^{\nu}
(\Omega_{\nu}^\ast (\tau_{\nu})/\Delta_{\nu})g_{\nu} \hat {A}_{\nu,o} (\tau_{\nu} ,Z)$, and assume  $\omega_{\nu}-\omega_{\nu}^c =\omega_{21}$. All put together, taking into account that $\hat {P}_{33,\nu}^j\cong 0$, and after adiabatic elimination of the excited atomic coherences $\hat {R}_{13,\nu}^j (t) \cong \frac{g_{\nu} \hat {A}_{\nu,o} (t,z_j )\hat {R}_{11,\nu}^j (t)+ \Omega
_{\nu}(t) \hat {R}_{12,\nu}^j (t )}{\Delta _{\nu} + \Delta _{31,\nu}^j  }$ and
$\hat {R}_{32,{\nu}}^j (t ) \cong \frac{g_{\nu} \hat {A}_{\nu,o}^ + (t,z_j )\hat
{R}_{12,{\nu}}^j + \Omega _{\nu}^\ast (t)\hat {R}_{22}^j (\tau )}{\Delta _{\nu} + \Delta
_{32,{\nu}}^j }$, we find:

\begin{eqnarray}
\begin{array}{ccc}
\label{Rabi field}
\textstyle{{\partial }\over{\partial Z}}\hat{\zeta}_{\nu} =
\textstyle{{i\beta _{\nu}}\over{2}}
\Big (\textstyle{{-(-1)^{\nu}} \over {\Delta_{\nu}}}
\hat B_{11,\nu}
\hat{\zeta}_{\nu} +
f_{\nu}
\hat B_{12,\nu}
 \Big ),\\
\label{R coherence2}
\textstyle{{\partial }\over{\partial \tau _{\nu}}}\hat {R}_{12,{\nu}}^j  =
-i(-1)^{\nu}
\hat{\zeta}_{\nu}
\textstyle{{\hat {R}_{11,\nu}^j  - \hat {R}_{22,\nu}^j}\over {1+\Delta_{31,\nu}^{j} / \Delta _{\nu}}}\\
- i\Big (\Delta _{R,tot,\nu}^j
-\textstyle{\Delta _{\nu} \over {1+\Delta_{31,\nu}^{j}/\Delta _{\nu}}}
(f_{\nu}-\textstyle{{\hat{\zeta}_{\nu}^{+}\hat{\zeta}_{\nu}}\over{|\Omega_{\nu}(\tau_{\nu})|^{2}}})\Big )
\hat {R}_{12,{\nu}}^j,\\
\textstyle{\partial \over {\partial \tau _{\nu}}}\hat {R}_{11,\nu}^j  =
-i(-1)^{\nu}
\Big (
\textstyle{{\hat{\zeta}_{\nu}^+\hat {R}_{12,{\nu}}^j }\over {1+\Delta_{31,\nu}^{j} / \Delta _{\nu}}}
-\textstyle{{\hat {R}_{21,{\nu}}^j \hat{\zeta}_{\nu}}\over {1+\Delta_{31,\nu}^{j} / \Delta _{\nu}}}
\Big ),
\end{array}
\end{eqnarray}

\noindent
where
$\textstyle{\partial \over {\partial t}}\hat {R}_{22,\nu}^j  = -
\textstyle{\partial \over {\partial t}}\hat {R}_{11,\nu}^j $  and $\beta_{\nu} =2\pi (n_o S)\vert
g_{\nu}\vert ^2/{v_{\nu}}$ with atomic density $n_{o}$, and cross section of the probe (echo)
fields $S$. $\hat B_{mn,\nu}=\int {d\Delta_{21,\nu}^j}\int {d\Delta _{31,\nu}^j}G(\Delta_{21,\nu}^j)$ $G(\Delta _{31,\nu}^j)\hat {R}_{mn,{\nu}}^j(\tau,Z_j\approx Z)/(1+\Delta_{31,\nu}^{j}/\Delta_{\nu})$, where $G(\Delta_{mn,\nu}^j)$ describes the IB of the $m\leftrightarrow n$ transition.

All variables in Eqs. (\ref{Rabi field}) depend on $\tau_\nu$ and $Z$. We emphasize that these equations hold for arbitrary numbers of atoms and photons. Compared to the usual Maxwell-Bloch equations \cite{Scully1997}, they contain additional non-linear terms, i.e. Stark shifts in the evolution  of the atomic coherence $\hat {R}_{12,\nu}^j $, and a population dependent term in the evolution of the light field $\hat{\zeta}_{\nu}$. Their analytic solution and analysis for time-reversal symmetry has not been considered before.

\emph{General reversibility of quantum dynamics:}
We now show that Eqs. (\ref{Rabi field}) allow for time reversible evolution
(i.e. storage and retrieval of the probe field) despite the non-linear terms. Indeed, the equations
for retrieval coincide with the equations for absorption for time-reversed echo emission (i.e. $\tau_2\rightarrow -\tau_2'$ and $\textstyle{{\partial} \over {\partial \tau _{2}}}\rightarrow-\textstyle{{\partial} \over {\partial \tau _{2'}}}$) if the following \emph{strong field conditions of reversibility} are satisfied:

\textit{i)} $c(\vec {K}_1 - \vec {K}_2 ) =(n_1 \omega_{1} +n_2\omega_{2})\vec{e}_z$, i.e. a phase (or mode) matching operation that results in mapping the atomic coherence created by the forward propagating probe field onto coherence that can create a backwards propagating echo field (i.e. ensures equal conditions at the end of the forward, and the beginning of the backward evolution). It is found from $\hat {R}_{12,2}^j (t^\prime )=\exp \{- i \alpha\} \hat{R}_{12,1}^j (t^\prime) \forall j$, where $t^\prime$ denotes a moment after complete probe absorption, and by expressing $\hat{R}_{12,1}^j$ and $\hat{R}_{12,2}^j$ through $\hat{P}_{12}^j$. $\alpha $ is a phase factor that contributes to the global phase of the echo signal. Note the absence of a similar equality for the light field operators, which is due to complete absorption of the probe field.

\textit{ii)} $\beta_1=\beta_2$ and $f_1(\tau_1)= f_2 (-\tau_2')$, i.e. equal coupling between the atomic coherence of the Raman transition and the probe and echo fields, resp., and $|\Omega_{1}(\tau_1)|=|\Omega_{2}(-\tau_2')|$, i.e. temporal reversibility of the Rabi frequencies of the writing and reading fields (see Fig.1).

\textit{iii)}
$\Delta_{R,tot,2}^{j}-
\textstyle{{f_{2}(-\tau_{2}')\Delta_{2}} \over {1+\Delta_{31,2}^{j}/\Delta_{2}}}
+\Delta_{R,tot,1}^{j}-\textstyle{{f_{1}(\tau_{1})\Delta_1}\over {1+\Delta_{31,1}^{j}/\Delta_{1}}}=\Delta_{R,tot,2}^{j}+\Delta_{R,tot,1}^{j}=0$, i.e. rephasing of atomic coherence when reversing the IB, similar to CRIB \cite{comment1}.
The first equal sign requires meeting conditions \textit{(ii)} and \textit{(iv)}.

\textit{iv)} $\Delta _{2} = - \Delta _{1}$, and
$\textstyle{{\Delta _{31,2}^j}\over{\Delta _{2}}} =
\textstyle{{\Delta _{31,1}^j}\over{\Delta _{1}}}$,
i.e. anti-correlated spectral detunings of the light fields. This condition completely determines $\omega_2$ for a given $\omega_1$ and $\Delta_1$: $\omega_{2}=\omega_{31}-\Delta_{2}=\omega_{1}+2\Delta_{1}$, where we used that $\omega_{31}=\omega_\nu+\Delta_\nu$.

We note that the time-reversibility hidden in Eqs. \ref{Rabi field} results in a temporally reversed replica of arbitrary probe fields in the echo field, which is equivalent to the standard, linearized system of Maxwell-Bloch equations \cite{Moiseev2001,Kraus2006}. Thus any quantum state encoded into the input light field can be stored and recalled with unit efficiency and fidelity, despite the nonlinear interactions. At the same time the demonstrated reversibility enables new possibilities for the realization of optical quantum memory, e.g. storage of macroscopic light fields in nano-size memories where the usual weak field approximation may not be satisfied.

\emph{Storage of weak probe fields:} It is interesting to compare the previous conditions for time-reversibility with the conditions in the case of weak probe fields,
where $\langle\hat{R}_{11,\nu}\rangle\cong1$ and $\langle\hat{R}_{22,\nu}\rangle\cong0$. In the following we also assume that the Stark shift due to the presence of the probe and echo fields is small compared to the spectral width of the stored light:
$\textstyle{{\Delta_{\nu}\langle\hat{\zeta}_{\nu}^{+}\hat{\zeta}_{\nu}\rangle}\over{|\Omega_{\nu}|^{2}}}\ll \delta\omega$ where $\langle ..\rangle$ denotes the expectation value,
$\textstyle{|\Delta_{31,\nu}^j|\ll\Delta_{\nu}}$, and we use new variables
$(\tilde{\hat{\zeta}}_{\nu},\tilde{\hat{R}}_{12,\nu})=(i\hat{\zeta}_{\nu},\hat{R}_{12,\nu})$
$exp \{i(-1)^{\nu}\frac{\beta _{\nu}}{2\Delta_{\nu}}Z-i\psi_{\nu}\}$ in Eqs. (\ref{Rabi field}), where $\psi_{\nu}=\int^{\tau_{\nu}}\Delta_{\nu}^s (\tau_\nu ) d\tau_{\nu}$. We find

\begin{eqnarray}
\begin{array}{ccc}
\label{weak field 1}
\textstyle{{\partial }\over{\partial Z}}\tilde{\hat{\zeta}}_{\nu} =
-\frac{\beta _{\nu} f_{\nu}}{2} \tilde{\hat{B}}_{12,{\nu}},
\\
\label{weak field 2}
\textstyle{{\partial }\over{\partial \tau _{\nu}}}\tilde{\hat{R}}_{12,{\nu}}^j  =
- i\Delta _{R,tot,\nu}^j
\tilde{\hat{R}}_{12,{\nu}}^j -(-1)^{\nu}\tilde{\hat{\zeta}}_{\nu},
\end{array}
\end{eqnarray}

\noindent
where the centers of the Raman transitions are shifted by $\Delta_{\nu}^s(\tau_\nu )=\Delta_{\nu} f_\nu (\tau_\nu )$ (see Fig.1).
 For a probe field with symmetric shape in time, and a Raman broadening that is symmetric in frequency, these linearized equations describe again time reversed evolution of the atom-light system, but this time under the \emph{weak field conditions of reversibility} \cite{comment2}:

\textit{i')} $c(\vec {K}_1 - \vec {K}_2 ) =
\vec {e}_z [ n_1\omega_{1} + n_2 \omega_{2}
+ c(  \beta_1/\Delta _{1}+\beta_2/\Delta _{2} )]$, which is found from $\tilde{\hat {R}}_{12,2}^j (t^\prime )=
\exp \{ -i\alpha '\} \tilde{\hat{R}}_{12,1}^j (t^\prime)$,
as explained in (\textit{i}).
Note that this generalized mode matching condition coincides with (\textit{i}) if conditions \textit{(ii)} and \textit{(iv)} are met. Yet, in the case of weak probe fields, these requirements are relaxed, e.g. due to the lack of the probe (echo) induced Stark shift. In particular, this allows for continuous frequency conversion of the echo compared to the probe.
Assuming resonance with the center of the IB Raman transition and constant Stark shifts, we find $\omega_1=\omega_1^s+\omega_{21}-\Delta_1^s$ and $\omega_2=\omega_1+\omega_2^c-\omega_1^c+\Delta_1^s-\Delta_2^s$.

\textit{ii')} $\beta_1 f_1(\tau_1)=\beta_2 f_2(-\tau_2)$, similar to \textit{ii)}, but without the necessity to individually equalize each variable.

\textit{iii')}
$\Delta _{R,tot,2}^j t_2 +
\Delta _{R,tot,1}^j t_1=k n_j 2\pi + const$, where $k\in \textbf{N}_{0}$, $n_j\in \textbf{Z}$, and $const$ is an irrelevant constant. We emphasize that this generalized rephasing condition can be met using additional experimental approaches compared to (\textit{iii}). We can find those approaches, characterized by $k$, by identifying situations where all atoms accumulated the same phase (modulo $2\pi$), i.e. where the initially excited collective coherence is recovered.

For $k=0$, rephasing is achieved at time $t_1=t_2$ by actively inverting the Raman detuning of each atom $j$, and we recover condition (\textit{iii}). In analogy to CRIB \cite{Kraus2006}, we refer to this protocol as RECRIB.

For $k=1$, we analyze Eqs. \ref{weak field 1} for a symmetric comb structure, where each atom $j$ is located in an absorption line detuned from the unbroadened transition by a multiple of $\delta_{comb}$ given by $n_j$: 
$\Delta_{31,tot,1}^j=\Delta_{31,tot,2}^j =n_j\delta_{comb}$ (we assume for simplicity that the broadening of the Raman transition is determined by the broadening of the $1\leftrightarrow 3$ transition and $f_\nu$). Using symmetry properties for absorption and retrieval to recover the coherence $\tilde{\hat{B}}_{12,{\nu}}$, we find an echo pulse emitted at a time given by $f_1t_1+f_2t_2=2\pi /\delta_{comb}$, which can be controlled by varying $\delta_{comb}$, $f_1$, or $f_2$. In analogy to the recently proposed AFC QM \cite{Afzelius2009}, we refer to this approach as REAFC. Approaches with $k>1$ are subclasses of $k\neq 0$. They also rely on atomic frequency combs, but exploit that rephasing of atomic coherence is repetitive, i.e. can generate echoes at later times.

\emph{Efficiency:} To find the efficiency $\epsilon=\int dt \langle\hat{A}^\dagger_2(t)\hat{A}_2(t)\rangle/$ $\int dt \langle\hat{A}^\dagger_1(t)\hat{A}_1(t)\rangle$ of REQM (a crucial property for a quantum repeater) in a realistic case, we have to consider limited absorption, and the Raman IB, as determined by the $1\leftrightarrow 3$ IB and $f_\nu$. Towards this end, we analytically solve Eqs. (\ref{weak field 1}). Assuming the case without
frequency conversion ($\beta _\nu\equiv\beta $,
$\upsilon _\nu\equiv\upsilon $, $\Delta _{\nu}\equiv\Delta$,
$f_\nu\equiv f$), and Gaussian line shapes, we find
\begin{equation}
\label{efficiency}
\epsilon = exp\{-\Gamma^2 (t_1+t_2)^2\}\vert 1 - e^{ -
\alpha_{eff} L}\vert ^2.
\end{equation}
$\alpha _{eff}L$ is the effective optical depth, which depends on the on-resonant absorption coefficient $\alpha_o=\beta \sqrt{\pi/2}/\Delta_{n,nat}^{(31)}$, $L$ is the length of the atomic medium, and $\Gamma$ is related to IB that leads to irreversible dephasing of atomic coherence. In the case of RECRIB, assuming that the width of the initial, naturally broadened absorption line $\Delta_{n,nat}^{(31)}$ (possibly reduced through spectral tailoring \cite{Alexander2006}), is small compared to the line width $\Delta_{n,cont}^{(31)}$ after controlled broadening, we find $\alpha _{eff}^{(RECRIB)}=\alpha_o \Delta_{n,nat}^{(31)}/\Delta_{n,cont}^{(31)}$ and $\Gamma^{(RECRIB)}=f\Delta_{n,nat}^{(31)}$. For REAFC, assuming that the width $\gamma$ of each individual line of the comb is small compared to the width of the whole comb structure, and that the comb has a large number of lines, we find  $\alpha _{eff}^{(REAFC)}=\alpha_o \sqrt{2\pi}(\gamma/\delta_{comb})$ and $\Gamma^{(REAFC)}=f\gamma$.

For RECRIB, we normalize all units to the width of the broadened Raman transition ($f\Delta_{n,cont}^{(31)}=1$), and we set the spectral width of the probe field to $\delta\omega\equiv 0.5$. This leads to $\delta t\equiv 2$ and $t_1=t_2\equiv 2 \delta t$, hence $t_1+t_2=8$. Note that these definitions for $t_1$ and $t_2$ describe storage of a single temporal mode.
For REAFC, we assume the width of the Raman comb structure $f\Delta_{comb}\gg \delta\omega\gg f\delta_{comb}\equiv 1$, i.e. we normalize the line spacing in the Raman comb structure to one (note the different normalization compared to RECRIB). This leads to $t_1+t_2=2\pi$.

Fig. 2 shows the efficiency for both protocols as a function of $\Gamma^{RECRIB}\in [0..1]$ and $\Gamma^{REAFC}\in [0..1]$, resp., for different on-resonant optical depths $\alpha_oL$. The general shapes of the efficiency curves are similar, yet, with REAFC reaching higher efficiency for a given depth. The optimum value for $\Gamma$ is determined by the tradeoff between small irreversible dephasing, i.e. small individual unbroadened absorption lines ($\Delta_{n,nat}^{(31)}$ and $\gamma$, resp.), and large absorption, i.e. large line width compared to $\Delta_{n,cont}^{(31)}$ or $\delta_{comb}$, resp. Furthermore, the maximum efficiency increases with $\alpha_oL$, and the difference between the two protocols decreases.

\begin{figure}
  \includegraphics[width=0.4\textwidth,height=0.2\textwidth]{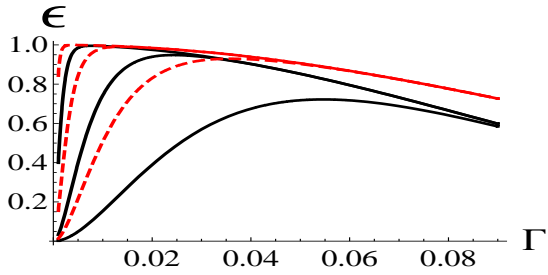}
  \caption{Quantum efficiency $\varepsilon$ for single-mode storage as a function of $\Gamma=(\Gamma^{RECRIB},\Gamma^{REAFC})$ for resonant optical depths $\alpha_o L = 50, 200,
1000$ (bottom to top sets of two traces). }
  \label{Figure2}
\end{figure}

\emph{Conclusion:} We have shown that REQM generalized time-reversibility in photon-echo quantum memory to storage of strong light fields and non-linear interactions. Furthermore, it unifies AFC and CRIB in an extension to off-resonant Raman transitions, and allows using atomic materials with short optical coherence times. It also allows exploiting additional degrees of freedom such as wave vectors, and carrier and Rabi frequencies of the control fields. This leads to the possibility to influence Raman IB, i.e. allow larger storage bandwidth, and to a large variety of manipulations of light fields, e.g. frequency conversion and generalized quantum compression \cite{Moiseev2008}. Note that RECRIB \cite{Hetet2008b} and AFC \cite{Riedmatten2008} have recently been demonstrated in atomic vapor and rare-earth-ion doped crystals, resp., hence REQM can be explored with present technology.

Financial support by the Natural Sciences and Engineering Research
Council of Canada, General Dynamics
Canada, Alberta's Informatics Circle of Research Excellence, and the Russian Fund for Basic Research grant \#
08-07-00449 is gratefully acknowledged.


\begin{references}
\bibitem{comment1} It may be possible to generalize this expression, similar to the below discussed case of weak probe fields.
\end{references}
\end{document}